\documentclass[a4paper]{spie}  %>>> use this instead for A4 paper
\usepackage[]{graphicx}
\usepackage{amssymb,amsmath,psfrag,url}
\usepackage{verbatim}

\newcommand{\Bolivarallee}{Boliva\hspace{-0.1mm}r\hspace{0.15mm}a\hspace{-0.1mm}llee}

\newcommand{\Takustrasse}{Taku\hspace{0.25mm}s\hspace{-0.1mm}tra{\ss}e}

\newcommand{\Field}[1]{{\bf{#1}}}
\newcommand{\Tensor}[1]{{\bf{#1}}}

\newcommand{\myset}[1]{\left\{#1\right\}}

\newcommand{\real}{\mathbb{R}}

\newcommand{\curl}{\mathbf{curl}\;}
\newcommand{\curlz}{\mathbf{curl}}

\newcommand{\hcurl}{\mathrm{H}\left(\curlz\right)}

\title{hp-finite-elements for simulating electromagnetic fields \\ in optical devices with rough textures}

\author{
Sven~Burger,\supit{\,ab}
Philipp~Gutsche,\supit{\,ab}
Martin~Hammerschmidt,\supit{\,b}
Sven~Herrmann,\supit{\,b}
Jan~Pomplun,\supit{\,a}
Frank~Schmidt,\supit{\,ab}
Benjamin~Wohlfeil,\supit{\,b}
Lin~Zschiedrich\supit{\,a}
\skiplinehalf
\supit{a}
JCMwave GmbH,
\Bolivarallee~22, 
D\,--\,14\,050 Berlin,
Germany
\smallskip\\
\supit{b}
Zuse Institute Berlin\,(ZIB),
\Takustrasse~7,
D\,--\,14\,195 Berlin,
Germany
\authorinfo{
Corresponding author: S.~Burger\\
URL: http://www.jcmwave.com\\
URL: http://www.zib.de
}}

\begin{document}
\maketitle
%%%%%%%%%%%%%%%%%%%%%%%%%%%%%%%%%%%%%%%%%%%%%%%%%%%%%%%%%%%%% 
%% SPIE Copyright form 
\noindent
This paper will be published in Proc.~SPIE Vol.~{\bf 9630}
(2015) 96300S ({\it Optical Systems Design 2015: Computational Optics}, DOI: 10.1117/12.2190119)
and is made available 
as an electronic preprint with permission of SPIE. 
One print or electronic copy may be made for personal use only. 
Systematic or multiple reproduction, distribution to multiple 
locations via electronic or other means, duplication of any 
material in this paper for a fee or for commercial purposes, 
or modification of the content of the paper are prohibited.
Please see original paper for images at higher resolution. 
%%%%%%%%%%%%%%%%%%%%%%%%%%%%%%%%%%%%%%%%%%%%%%%%%%%%%%%%%%%%% 

\begin{abstract}
The finite-element method is a preferred numerical method when electromagnetic fields at high accuracy 
are to be computed in nano-optics design.  
Here, we demonstrate a finite-element method using {\it hp}-adaptivity on tetrahedral meshes for 
computation of electromagnetic fields in  a device with rough textures.
The method allows for efficient computations on meshes with strong variations in element sizes. 
This enables to use precise geometry resolution of the rough textures. 
Convergence to highly accurate results is observed.

\end{abstract}


\section{Introduction}

Optical instruments and experimental setups modify and detect 
light fields in various spatial, spectral and temporal dimensions. 
In many state-of-the-art systems components with sub-wavelength 
geometry features are included, utilizing, e.g., interference effects of 
the electromagnetic field. 
The large design space where many parameters can be tuned 
allows to obtain specific effects and to optimize instruments for 
dedicated functionalities. 
Rigorous Maxwell solvers are required in the design flow typically when 
design on a nanometer scale impacts system performance quantitatively. 
Various numerical methods are used to solve Maxwell's
equations rigorously~\cite{Taflove2005computational,Jin2014finite,Lavrinenko2014crc}.
A main challenge for Maxwell solvers 
is typically efficiency (i.e., to achieve accurate results at low computation times). 
Finite-element methods (FEM) allow for high efficiency due to accurate geometry modelling, 
adaptive meshing strategies, and higher-order convergence. 
In simulation tasks requiring high accuracy FEM can outperform other rigorous simulation 
methods~\cite{Maes2013oe,Hoffmann2009spie}. 

\begin{figure}[b]
\begin{center}
%\psfrag{dRs}{\sffamily $dR_\textrm{S}$}
\includegraphics[width=.95\textwidth]{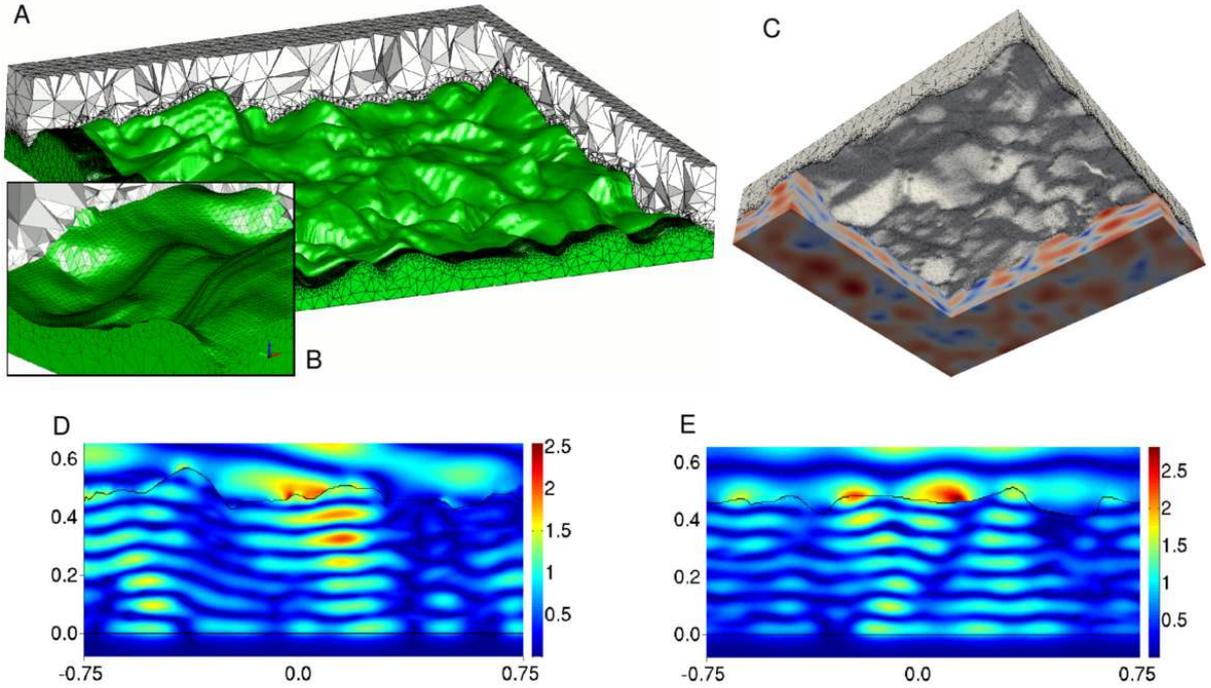}
%\hspace{5mm}
  \caption{
{\bf A}: Visualization of a geometry with a rough interface with parts of the geometry discretization (tetrahedral mesh).  
{\bf B}: Detail of A, showing small mesh elements at the interface.
{\bf C}: Visualization of a computed light intensity distribution in the 3D device (view from below). 
{\bf D}: Visualization of the light intensity distribution in a x-z-cross section through the 3D device. 
{\bf E}: Visualization of the light intensity distribution in a y-z-cross section. 
}
\label{fig1}
\end{center}
\end{figure}

We develop and investigate finite-element methods for electromagnetic field simulations. 
Recently, our FEM Maxwell solver ({\it JCMsuite}) has been applied for various tasks in optics design.
Among these are the design of 
microstructured fibers~\cite{Bock2013lpr,wong2012excitation}, 
solar cells~\cite{paetzold2011design}, 
microlithography masks~\cite{Tyminski2013al},
integrated microlenses~\cite{Gschrey2015nc}, 
metamaterials~\cite{Kaschke2014oe},  
microcavities~\cite{Shchukin2014ieee,Karl2009OE}, 
plasmonic waveguides~\cite{Kewes2013design},
scatterometry reference standards~\cite{Bodermann2012op},
and others~\cite{Burger2013pw}.

Here, we discuss a method for improvement of numerical performance, so-called {\it hp}-FEM. 
We apply the method 
for efficient simulation of 3D devices with rough surface textures. 
Figure~\ref{fig1} shows the model, mesh, and simulation result of a typical investigated sample. 
In particular, the usage of {\it hp}-FEM allows for very fine geometrical resolution of the surface texture at only moderately
increased computational cost. 

This paper is structured as follows: 
The background of the FEM and {\it hp}-FEM is presented in Section~\ref{section_background}. 
The method is validated in Section~\ref{section_application}, 
by presenting simulation results for an example of a rough texture  
related to requirements in thin-film solar cell design. 

%%%%%%%%%%%%%%%%%%%%%%%%%%%%%%%%%%%%%%%%%%%%%%%%%%%%%%%%%%%%%%%%%%%%%%%%%%%%%%
% HP INTRODUCTION
%%%%%%%%%%%%%%%%%%%%%%%%%%%%%%%%%%%%%%%%%%%%%%%%%%%%%%%%%%%%%%%%%%%%%%%%%%%%%%

\section{Hp finite-element method}
\label{section_background}
In the following the background of the finite-element method is summarized~\cite{MON03,Babuska1981}, 
following a previous notation~\cite{Pomplun2007pssb,Burger2015al}.
Light scattering off nanoscopic structures is modeled by 
the linear Maxwell's equations in frequency domain. 
From these a single equation for the electric field $\Field{E}$ can be derived:
\begin{equation}
  \curl\Tensor{\mu}^{-1}\curl \Field{E}-\omega^{2}\Tensor{\epsilon}\Field{E}=i\omega\Field{J},
  \label{eq:mwE}
\end{equation}
where  $\Tensor{\epsilon}$ and $\Tensor{\mu}$ are the permittivity and permeability tensor, $\omega$ is 
the time-harmonic frequency of the electromagnetic field, and the  
electric current $\Field{J}$ is a source of an electromagnetic field. 
The domain of interest is separated into an infinite 
exterior $\Omega_{\mathrm{ext}}$ which hosts the given incident field and the scattered field, 
and an interior $\Omega_{\mathrm{int}}$ where the total field is computed. 
Electromagnetic waves incident from the exterior to the interior at the boundaries between both domains
are added to the right hand side of Eq.~\eqref{eq:mwE}. 
For numerical simulations the infinite exterior is treated using transparent 
boundary conditions (using the perfectly matched layer method, PML).

For a FEM discretization,  Eq.~\eqref{eq:mwE} is first transformed into a weak formulation, i.e., 
it is tested with a vectorial function $\phi$ and integrated over $\real^{3}$ which yields:

\begin{equation}
\int\limits_{\real^{3}}(\curl \phi) \;\Tensor{\mu}^{-1}\curl \Field{E}-\omega^{2}\phi\,\Tensor{\varepsilon}\Field{E}
=
i\omega\int\limits_{\real^{3}}\phi\,\Field{J}.
   \label{eq:1}
\end{equation}

For compact notation, the forms $a(\phi,\Field{E})$ and $f(\phi)$ are introduced, 
and 
the function space $\hcurl$ is defined.
The weak form of Maxwell's equations then reads:
\\\noindent
Find $\Field{E}\in\hcurl$ such that:
\begin{equation}
  \label{eq:mwEweak}
a(\phi,\Field{E})=f(\phi)\,,\quad\forall \phi\in\hcurl.
\end{equation}

A finite-element discretization of Maxwell's equations restricts the 
formulation \eqref{eq:mwEweak} to a finite-dimensional subspace $V_h$ with $\dim V_h=N<\infty$:\\\noindent
Find $\Field{E}_h\in V_h$ such that:
\begin{equation}
  \label{eq:mwEweakFEM}
a(\phi_h,\Field{E}_h)=f(\phi_h)\,,\quad\forall \phi_h\in V_h.
\end{equation}

Next, a basis $\myset{\varphi_1,\dots,\varphi_N}$ 
of $V_h$ is constructed, and the electric field is expanded using the basis elements: $\Field{E}_h=\sum\limits_{i=1}^N e_i \varphi_i$. 
The variational problem (\ref{eq:mwEweakFEM}) is then tested with all elements of the basis which gives a linear system of equations:
\begin{equation}
\sum\limits_{i=1}^N a(\varphi_j,\varphi_i)e_i=f(\varphi_j)\,,\quad\forall j=1,\dots,N.
\label{equation_matrix}
\end{equation}
The matrix $A_{ji}=a(\varphi_j,\varphi_i)$ is sparse and can be decomposed with efficient sparse 
LU solvers to obtain the unknown expansion coefficients $e_i$ of the electric field.

The basis $\myset{\varphi_1,\dots,\varphi_N}$  is constructed using elements $\varphi_i$ (also called {\it ansatz functions}) 
which are polynomial functions of 
order $p$ (also called {\it finite-element degree}), 
and which are defined on a single patch of the spatial discretization of the geometry (mesh) only. 
For the results presented here, we attribute elements of different polynomial order $p$ to different patches of the 
mesh. 
In regions where the mesh is very fine (small $h$) due to required geometry resolution (fine details of the rough interfaces or very thin layers)
a lower polynomial order $p$ can be chosen than in regions where the mesh is coarser (large $h$). 
The method to distribute different orders $p$ to the different patches relies on estimating errors on the different patches, making use of  
informations on geometry, meshing, material properties and source fields~\cite{Burger2015al}. 
This yields a basis $\myset{\varphi_1,\dots,\varphi_N}$ which is well adapted to the problem and does not need too much computational 
effort in regions where it is not required.

%%%%%%%%%%%%%%%%%%%%%%%%%%%%%%%%%%%%%%%%%%%%%%%%%%%%%%%%%%%%%%%%%%%%%%%%%%%%%%
% ROUGH INTERFACE
%%%%%%%%%%%%%%%%%%%%%%%%%%%%%%%%%%%%%%%%%%%%%%%%%%%%%%%%%%%%%%%%%%%%%%%%%%%%%%

%%%%%%%%%%%%%%%%%%%%%%%%%%%%%%%%%%%%%%%%%%%%%%%%%%%%%%%%%%%%%%%%%%%%%%%%%%%%%%
\section{Simulation of light propagation through thin films\\ with 3D rough interfaces}
\label{section_application}

\begin{figure}[b]
\begin{center}
\psfrag{x}{\sffamily x} 
\psfrag{y}{\sffamily y} 
\psfrag{z}{\sffamily z} 
\psfrag{px}{\sffamily $p_\textrm{x}$} 
\psfrag{py}{\sffamily $p_\textrm{y}$} 
\psfrag{h1}{\sffamily $h_\textrm{r}$} 
\psfrag{h2}{\hspace{-3mm}\sffamily $h_\textrm{aSi}$} 
\psfrag{TCO}{\sffamily TCO} 
\psfrag{Si}{\sffamily a-Si} 
\psfrag{metal}{\sffamily Ag} 
\includegraphics[width=.5\textwidth]{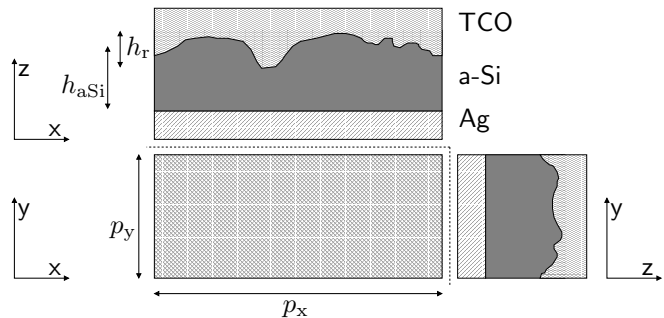}
%\hspace{5mm}
  \caption{
Schematic of the investigated setup. 
}
\label{fig2}
\end{center}
\end{figure}

In this section we demonstrate the performance of {\it hp}-FEM for an application related to photovoltaics.
A design task in thin-film photovoltaics is light trapping~\cite{Battaglia2012acsn}: 
In order to avoid reflection of light from solar cells and to reach high absorption of 
light in relatively thin layers of moderate absorptance, interfaces are equipped with 
rough (irregular) or with regular (e.g., periodic) nano textures. 
In order to design the device for best efficiency (maximum absorption), typically 
Maxwell's equations have to be solved for the 3D device. 
Geometrical design parameters are layer thicknesses of the material stacks as well as 
various accessible dimensions of the nano textures.   
Depending on the computational domain size this can be computationally demanding. 
For regular nano textures, typically it is sufficient to use a single 
unit cell of a periodic array as computational domain. 
However, for rough textures larger computational domains have to be used which can significantly 
increase computational costs. 
We use {\it hp}-FEM for simulation of light propagation and 
scattering from rough interfaces
and aim for improved numerical performance for this type of setups. 

\begin{table}
\begin{center}
\begin{tabular}{|l|l|}
\hline
%material & Si \\ \hline
$p_x=p_y$ & [250 \dots 2000]\,nm\\ \hline
$h_\textrm{aSi}$ & 500\,nm\\ \hline
$h_\textrm{r}$ & 200\,nm\\ \hline
$\epsilon_{aSi}$ & 18.15+ 0.37i \\ \hline
$\epsilon_{Ag}$ & -17.03 + 1.15i \\ \hline
$\epsilon_{TCO}$ & 3.18 \\ \hline
$\lambda_0$& 650\,nm \\ \hline
\end{tabular}
\caption{Parameter settings (compare Fig.~\ref{fig2}).
}
\label{tab1}
\end{center}
\end{table}

\subsection{Model}
\label{sec_model}
As model system we investigate a material stack composed of a metal layer (back reflector), 
a thin amorphous silicon (aSi) layer and a transparent conductive oxide (TCO: SnO$_2$).
Figure~\ref{fig2} shows a schematic of the setup. 
As indicated, the interface between aSi and TCO is rough while
the interface between metal and aSi is assumed to be flat. 
To model the roughness we use topography data which can be obtained, e.g.,  using atomic force microscopy. 
The roughness data typically is given as $z$-positions on a regular, Cartesian grid ($x, y$-coordinates). 
We perform pre-processing in Matlab on this data in order to periodify it. This also allows to possibly
coarsen too fine meshes or to locally refine meshes around specific features. 
A surface mesh discretizing the rough topology is then passed to the tetrahedral mesh generator included in 
our FEM solver JCMsuite. 
Figure~\ref{fig1} shows a typical mesh resulting from this procedure. 
Please note that the mesh elements adjacent to the rough interface are very small while 
other mesh elements are significantly larger. 
The mesh element volumes differ by several orders of magnitude. 
This allows to accurately model the complex geometry of the rough surface while restricting the 
number of mesh elements to a moderate number. 
Further, a domain-decomposition method is used to efficiently treat light propagation in unstructured domains
of the geometry~\cite{Schaedle_jcp_2007}. 

Geometry and material parameters of the model are summarized in Table~\ref{tab1}.
The optical properties (permittivity) of the involved materials at the 
given wavelength $\lambda_0$ are obtained from the literature~\cite{Ding2011semsc,Palik1985}.
As illumination we choose exemplary a $P$-polarized plane wave incident from superspace at an inclination angle of $\theta=10\,$deg 
w.r.t.~the $-z$-direction.

\subsection{Convergence results}
The quantity of interest in the investigated application is the amount of light absorbed in the 
aSi-layer. 
We compute this by integrating over the electromagnetic field energy density in the aSi domain. 
The imaginary part of this quantity is related to the absorbed field energy. 
We deduce the absorption in this layer, $A_\textrm{aSi}$, by relating the absorbed field energy to the field energy flux density 
of the incident plane wave, integrated over the computational domain boundary~\cite{Lockau2011rigorous}. 

\begin{figure}[b]
\begin{center}
\psfrag{n}{\sffamily N} 
\psfrag{Rel. Err.}{\hspace{-0.5cm}\sffamily $\Delta_\textrm{rel} (A_\textrm{aSi})$} 
\psfrag{p}{\sffamily $p$} 
\psfrag{Pprec}{\sffamily $p_\textrm{Prec}$} 
\psfrag{N [1e6]}{\hspace{-0.9cm} \sffamily $N_\textrm{unknowns}[10^6]$} 
%\psfrag{p-refinement}{\tiny \sffamily $p$-refinement} 
%\psfrag{hp-refinement}{\tiny \sffamily $hp$-refinement} 
\psfrag{p-refinement}{\scriptsize \sffamily $p$-refinement} 
\psfrag{hp-refinement}{\scriptsize \sffamily $hp$-refinement} 
\includegraphics[width=.32\textwidth]{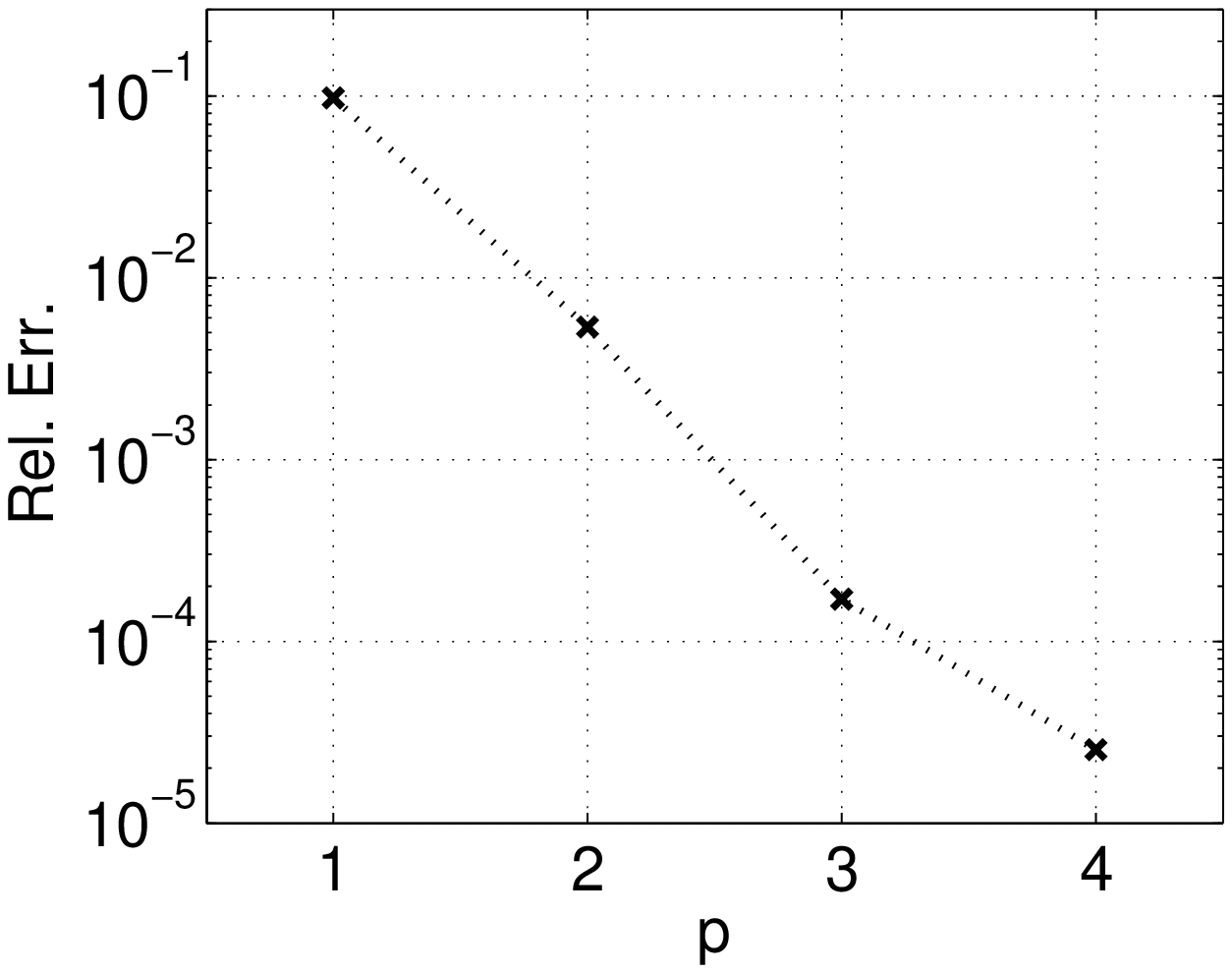}
\includegraphics[width=.32\textwidth]{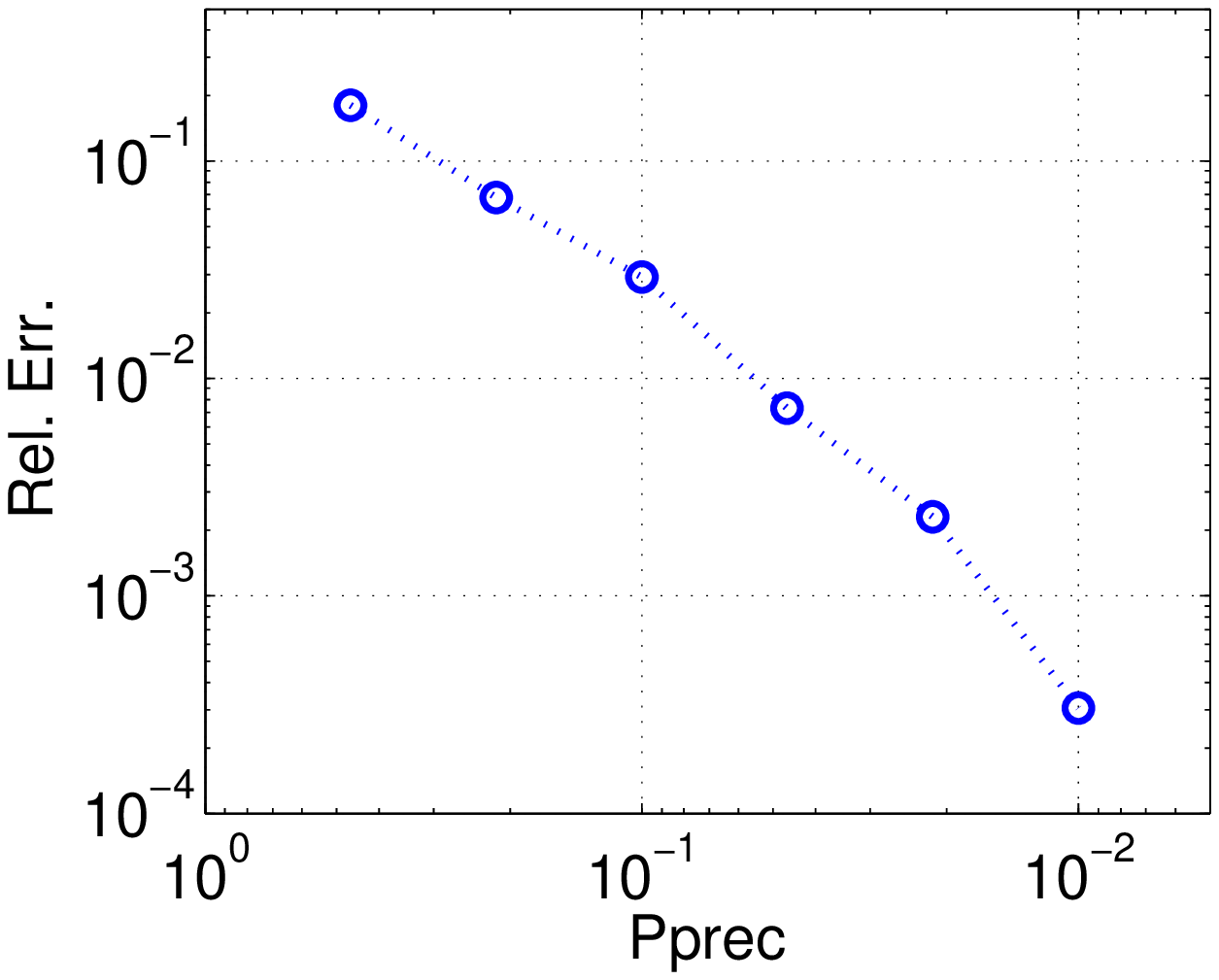}
\includegraphics[width=.32\textwidth]{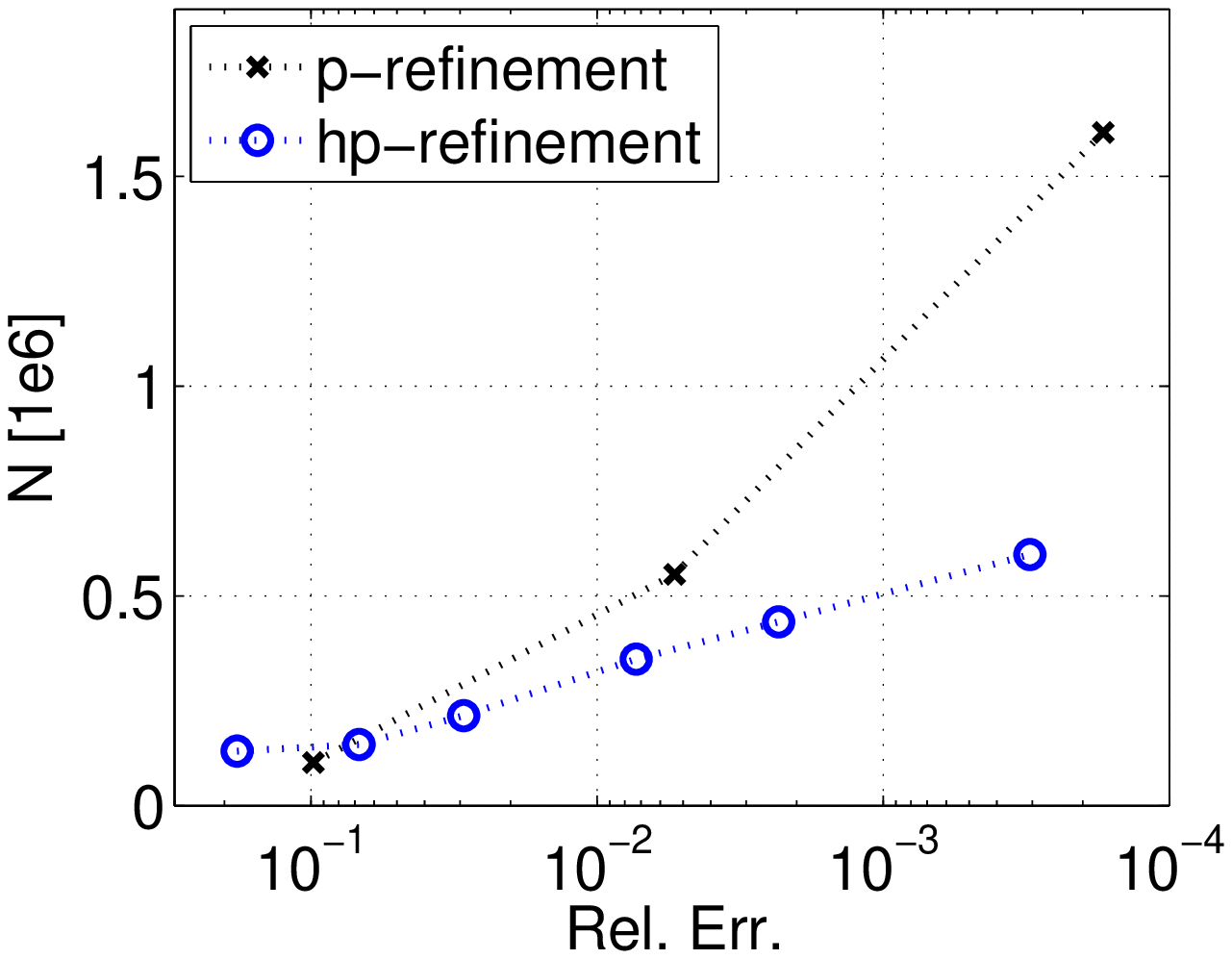}
%\hspace{5mm}
  \caption{
{\it Left:} Convergence ({\it p}-FEM) of the relative error of absorption in the aSi layer, $\Delta_\textrm{rel} (A_\textrm{aSi})$,
 with finite-element degree $p$. 
{\it Center:} Convergence ({\it hp}-FEM) with precision parameter $p_\textrm{Prec}$. 
{\it Right:} Number of unknowns of the discrete problem as a function of obtained accuracy for 
$p$-FEM and $hp$-FEM. 
}
\label{fig3}
\end{center}
\end{figure}

\subsubsection{p-convergence}
\label{sec_p_conv}
For the given model (as defined in Section~\ref{sec_model} with $p_x=p_y=500\,$nm) 
we compute the absorption $A_\textrm{aSi}$ at different numerical resolutions. 
In a first convergence scan we use the same finite-element degree $p$ on all mesh elements in each computation. 
As expected, we achieve higher accuracies with increasing $p$. 
Figure~\ref{fig3} (left) shows how the relative error of absorption, $\Delta_\textrm{rel} (A_\textrm{aSi})$, converges with $p$. 
The relative error is defined as relative deviation from the so-called 
{\it quasi-exact} result, $A_\textrm{aSi,qe}$, %Where the quasi-exact result, $A_\textrm{aSi,qe}$, is 
obtained using the same model and even higher 
finite-element degree ($p=5$). 
Hence, $\Delta_\textrm{rel} (A_\textrm{aSi}) = |A_\textrm{aSi}-A_\textrm{aSi,qe}|/A_\textrm{aSi,qe}$.
This is justified for setups where the asymptotic convergence regime is reached. 
It can be expected that this is the case here, because exponential convergence with $p$ (as seen in Fig.~\ref{fig3}, left) would typically 
not be obtained otherwise. 
%I.e.
Figure~\ref{fig3} (left) shows that when increasing $p$ from 1 to 4 the relative error decreases roughly exponentially
by several orders of magnitude to an error of about $2\times 10^{-5}$. 
At the same time the number of unknowns of the discrete problem increases from about $1\times 10^{5}$ (for $p=1$)
to about  $7\times 10^{6}$ (for $p=4$, not shown in Fig.~\ref{fig3}, right). 

\subsubsection{hp-convergence}
\label{sec_hp_conv}
In a second convergence scan for the same model we use different finite-element degrees $p$ on the different patches 
of the mesh elements. For this investigation we allow $p$ to take the values $p=1, 2, \textrm{or~} 3$. 
The distribution of $p$ to the different patches is managed by an {\it a-priori} error estimator which is 
controlled by a single parameter, $p_\textrm{Prec}$ ({\it precision parameter})~\cite{Burger2015al}. 
When the precision parameter is set to a small number, e.g., $p_\textrm{Prec}= 10^{-2}$ (high precision)
more higher-order finite-elements will be used than in the case of higher $p_\textrm{Prec}$. 
Figure~\ref{fig3} (center) shows how the relative error of the absorption converges with this parameter. 
As expected, we observe convergence of the quantity of interest. 
The relative error in this case is again computed as the relative deviation from the same {\it quasi-exact} result 
as in Sec.~\ref{sec_p_conv}, obtained with global setting of $p=5$.
This demonstrates that the precision parameter allows to reliably control numerical accuracy in our {\it hp}-FEM method. 

Figure~\ref{fig3} (right) shows how the computational effort (number of unknowns, $N_\textrm{unknowns}$, i.e, dimension of the 
discrete system) relates to the obtained numerical accuracies of the quantity of interest ($\Delta_\textrm{rel} (A_\textrm{aSi})$), 
for the data displayed in Fig.~\ref{fig3} (left \& center).
Memory consumption and computation time increase roughly linear with $N_\textrm{unknowns}$ in the investigated regime. 
Comparing the data obtained using {\it hp}-refinement to the data obtained using p-refinement, a significant difference in 
computational effort can be observed. 
E.g., for reaching an accuracy of $\Delta_\textrm{rel} (A_\textrm{aSi})\approx 10^{-3}$ $hp$-FEM reduces the computational 
effort in terms of number of unkowns roughly by a factor of three.

\begin{figure}[b]
\begin{center}
\psfrag{N [1e6]}{\hspace{-0.9cm} \sffamily $N_\textrm{unknowns}[10^6]$} 
\psfrag{absorption}{\sffamily $A_\textrm{aSi}$} 
\psfrag{pitch [um]}{\sffamily $p_x [\mu m]$} 
\psfrag{sqr(pitch) [um2]}{\hspace{0.2cm}\sffamily $p_x^2 [\mu m^2]$} 
\psfrag{z}{\sffamily z} 
\includegraphics[width=.32\textwidth]{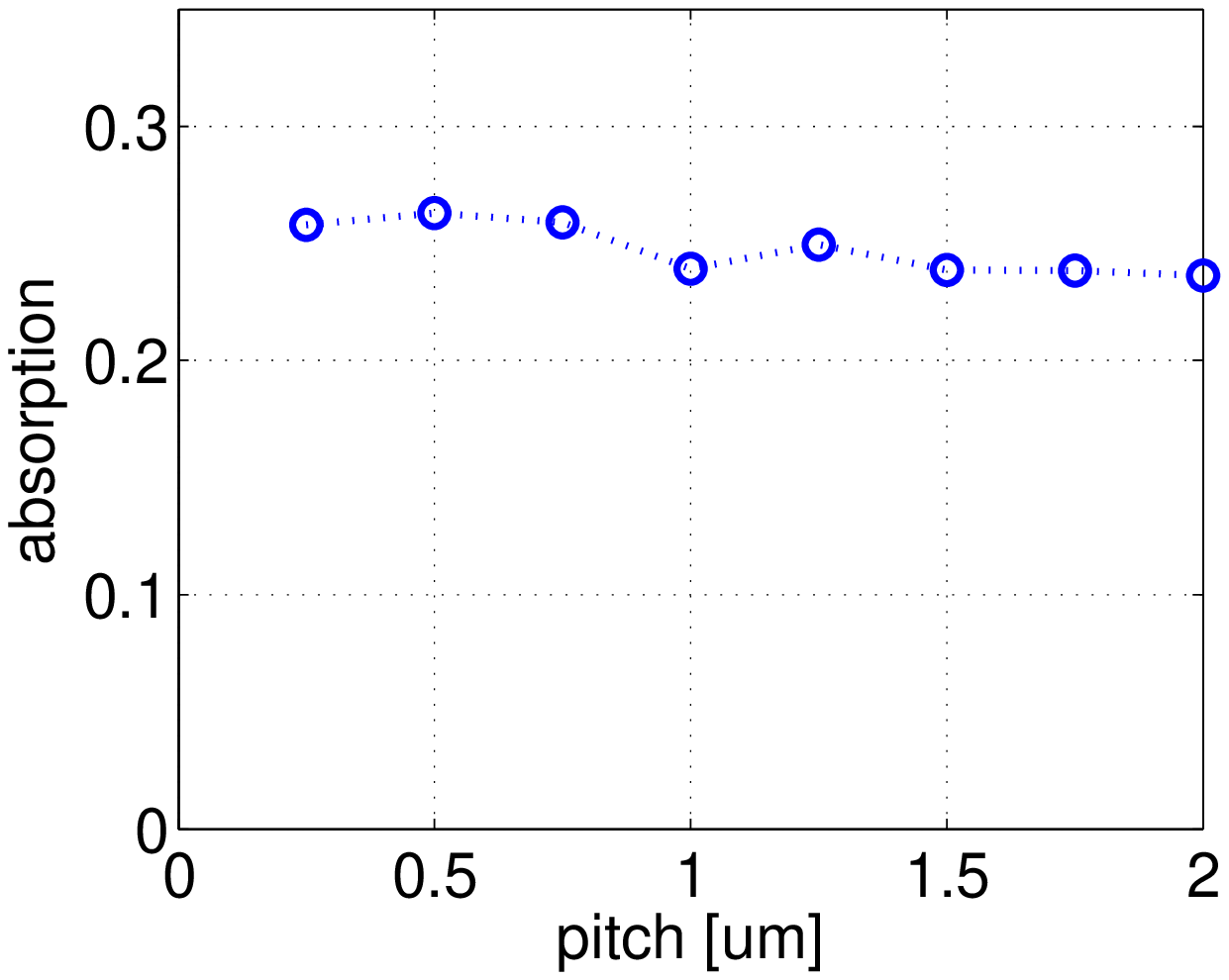}
\hspace{1cm}
\includegraphics[width=.32\textwidth]{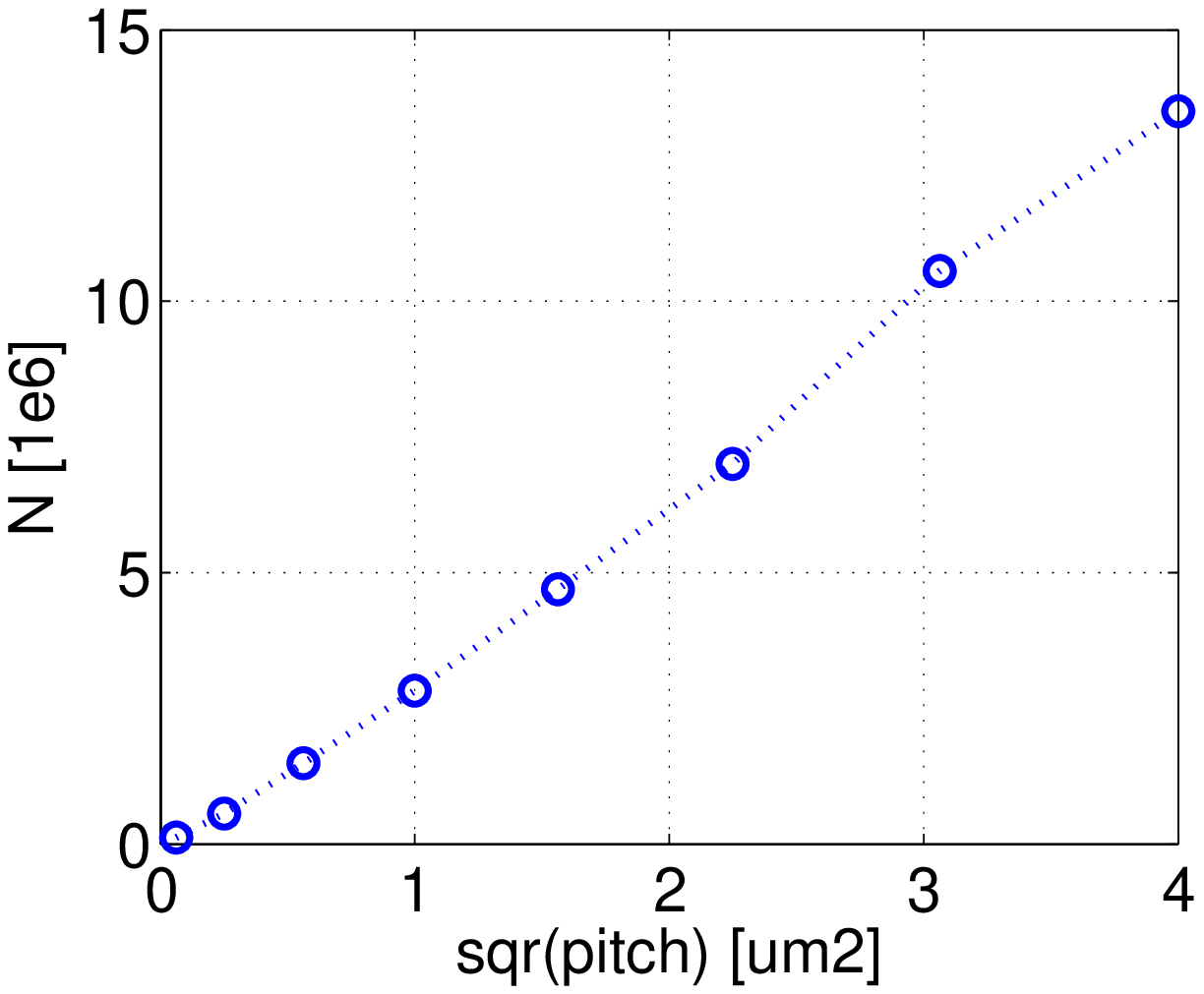}
%\hspace{5mm}
  \caption{
{\it Left:} Absorption in the rough aSi layer $A_\textrm{aSi}$ for varied computational domain size ($p_x=p_y$).
{\it Right:} Dependence of computational effort (number of unknowns, $N_\textrm{unknowns}$) on computational domain size. 
}
\label{fig4}
\end{center}
\end{figure}

\subsection{Performance for varied 3D computational domain size}
To demonstrate performance of the method we have varied the computational domain size and computed the absorption in the 
aSi layer, $A_\textrm{aSi}$. 
For these computations we have used {\it hp}-FEM with an accuracy setting $p_\textrm{Prec}= 10^{-2}$ ({\it cf.}\ Sec.~\ref{sec_hp_conv}).
Figure~\ref{fig4} (left) shows how the absorption at the given wavelength (see Table~\ref{tab1}) 
depends on the choosen computational domain size. 
Please note that due to the randomness of the rough interface the model is different for each computational domain size. 
However, one also can expect for a random interface that with increasing computational domain size the absorption should 
converge to some finite value~\cite{Wiesendanger2014apl}. 
For the data investigated in this study $A_\textrm{aSi}$ varies by less than 0.25\% for computational domains with 
$p_x \ge 1.5\,\mu$m. 
Figure~\ref{fig4} (right) displays the numerical effort (number of unknowns) for the respective computations. 
As expected, the number of unknowns depends roughly linear on the computational domain volume (which at constant height
is proportional to 
$p_x^2$). 

%%%%%%%%%%%%%%%%%%%%%%%%%%%%%%%%%%%%%%%%%%%%%%%%%%%%%%%%%%%%%%%%%%%%%%%%%%%%%%
% CONCLUSION ETC
%%%%%%%%%%%%%%%%%%%%%%%%%%%%%%%%%%%%%%%%%%%%%%%%%%%%%%%%%%%%%%%%%%%%%%%%%%%%%%

%%%%%%%%%%%%%%%%%%%%%%%%%%%%%%%%%%%%%%%%%%%%%%%%%%%%%%%%%%%%%%%%%%%%%%%%%%%%%%

\section{Conclusion}
A finite-element method using {\it hp}-adaptivity on tetrahedral meshes has been demonstrated for 
computation of electromagnetic fields and derived quantities in  a device with rough textures. 
Convergence to highly accurate results has been observed. 
The method allows to efficiently compute optical properties. % 
Fine, unstructured 
meshes can be choosen for accurate geometry discretization in specific regions of interest (in this case at the rough interface). 
In other regions where the geometry is simpler coarser meshes can be used. 
{\it hp}-FEM on such meshes with a variety of element sizes 
allows to attribute higher-order finite-elements (with higher computational costs) to large mesh elements 
and lower-order finite-elements (with lower computational costs) to small mesh elements. 
This results in relatively low computational costs for highly accurate FEM solutions. 
In essence, {\it hp}-FEM allows to treat problem setups with highly complex geometries on a nanometer (deep sub-wavelength) lengthscale 
while preserving the advantages of higher-order FEM convergence. 

\section*{Acknowledgments}
We acknowledge the support of BMBF through 
projects 13N13164 (SolarNano) and 13N12438 (MOSAIC)
and of the Einstein Foundation Berlin through projects ECMath\,-\,OT5 and -\,SE6.

\bibliography{osd15_sb_arxiv}
\bibliographystyle{spiebib}  

\end{document}